\documentclass[proceedings, preprint]{rmaa}

\usepackage{times}
\usepackage{graphicx}%,bm}%%,amssymb}
\usepackage{ulem}
\usepackage{epsfig}
\usepackage{natbib}
\usepackage{psfrag,color}

\def\prd{Phys.~Rev.~D}%% Physical Review D
\def\jcap{J. Cosmology Astropart. Phys.}%% Journal of Cosmology and Astroparticle Physics
\def\be{\begin{eqnarray}}
\def\ee{\end{eqnarray}}
\def\bi{\begin{itemize}}
\def\ei{\end{itemize}}
\def\lsim{\mathrel{\rlap{\lower3pt\hbox{\hskip1pt$\sim$}}
     \raise1pt\hbox{$<$}}} %less than or approx. symbol
\def\gsim{\mathrel{\rlap{\lower3pt\hbox{\hskip1pt$\sim$}}
     \raise1pt\hbox{$>$}}} %greater than or approx. symbol

\newcommand{\bary}{\begin{eqnarray}}
\newcommand{\eary}{\end{eqnarray}}

\SetYear{2013}
\SetConfTitle{Magnetic Fields in the Universe 4}

\title{Neutrino Oscillation from Hidden GRB Jets} 

\author{
  Nissim Fraija\altaffilmark{1} 
  and Enrique Moreno M\'endez\altaffilmark{2}}

\altaffiltext{1}{Luc Binette-Fundaci\'on UNAM Fellow. Instituto de Astronom\'\i{}a,  UNAM, CU, M\'exico (nifraija@astro.unam.mx).}
\altaffiltext{2}{Instituto de Astronom\'\i{}a,  UNAM, CU, M\'exico (enriquemm@astro.unam.mx).}

\shortauthor{N. Fraija, \& E. Moreno M\'endez}
\shorttitle{RevMexAA(SC) Demo Document}

\listofauthors{N. Fraija \& L. Author}
\indexauthor{N. Fraija}
\indexauthor{E. Moreno M\'endez}

\abstract{Collapsars are the likely progenitors of Long GRBs (lGRBs). 
lGRBs have been observed to last for thousands to tens of thousands of seconds, thus making unlikely
the neutrino-driven engine as the main mechanism for driving the jets. 
In this context, the Blandford-Znajek mechanism seems likely to explain the production of
rotational-axis directed jets without the need for large accretion rates. 
These engines, require magnetic fields between 10$^{12} \rm{G} < B < 10^{15}$ G threading the innerdisk,
Kerr-BH region to exist.  
We derive the neutrino self-energy and the effective potential up to O(1/$M_W^4$) in a weakly and highly magnetized GRB fireball flow which is made up of electrons, protons, neutrons and their anti-particles.  
We consider neutrino energies of 1-100 MeV which are produced during stellar collapse, merger events or in the fireball itself by electron-positron annihilation, inverse beta decay and nucleonic bremsstrahlung processes. Many of these neutrinos propagate through the fireball and may oscillate resonantly.
Using two-neutrino mixing  we study the possibility of these oscillations.}

\resumen{Los Col\'apsares son probablemente la fuente de los destellos de rayos gama de larga duraci\'on (lGRBs por sus siglas en ingl\'es).
Se han observado lGRBs con duraci\'on de miles y hasta decenas de miles de segundos, lo que hace poco probable que sus jets sean producidos por motores centrales accionados por neutrinos.
En \'este contexto, el mecanismo de Blandford-Znajek es un buen candidato para explicar la producci\'on de jets a lo largo del eje de rotaci\'on sin la necesidad de grandes tazas de accreci\'on.
Estos motores centrales requieren campos magn\'eticos de 10$^{12} \rm{G} < B < 10^{15}$ G atravezando la parte interna del dicso de accreci\'on y el agujero negro de Kerr.
Obtenemos la auto-energ\'ia y el potencial efectivo a orden O(1/$M_W^4$) para una bola de fuego hecha de electrones, protones, neutrones y sus antipart\'iculas respectivas y con campo magn\'etico d\'ebil y fuerte .
Consideramos que la energ\'ia de los neutrinos producidos durante el colapso gravitacional, as\'i como en fusi\'on de objetos compactos o en la bola de fuego en s\'i, por procesos de aniquilaci\'on de pares, decaimiento beta y por procesos bremsstrahlung nucle\'onicos es de entre 1 y 100 MeV.
Muchos de dichos neutrinos se propagan a trav\'es de la bola de fuego y pueden oscilar resonantemente.
Usando la mezcla de dos neutrinos  estudiamos la oscilaci\'on de neutrinos.
}

\addkeyword{binaries: general}
\addkeyword{black hole physics}
\addkeyword{gamma-ray bursts: general}
\addkeyword{stars: evolution}
\addkeyword{stars: magnetic field}
\addkeyword{stars: massive}

\begin{document}
% Typeset article header
\maketitle

Collapsars \citep{1993ApJ...405..273W,1999ApJ...524..262M} are generally accepted as the progenitors of Long Gamma-Ray Bursts (lGRBs).
This is supported by the prediction that lGRBs are accompanied by an energetic supernova (SN).
There is strong evidence that six lGRBs are spectroscopically associated with SNe (GRB980425/SN 1998bw, GRB 030329/SN 2003dh, GRB 031203/SN 2003lw, GRB 060218/SN 2006aj and GRB 100316D/SN 2010bh, see~\citealp{2011arXiv1104.2274H}; and GRB120422/SN 2012bz, see~\citealp{2012CBET.3100....2M}).  
There is also evidence of such a relation in another two dozen events.

One of the major issues with the Collapsar model, is the issue of producing a massive stellar core with enough angular momentum \citep[see, e.g.,][]{2005ApJ...626..350H}.  To solve this issue, one of the proposed mechanisms is to use a binary to transfer angular momentum back to the star late in its evolution \citep[see, e.g.][]{1998ApJ...494L..45P} .\\
\citet{2002ApJ...575..996L} studied several Galactic black-hole binaries (BHBs) and reproduced the natal spin of the BHs.
Using these results, \citet{2007ApJ...671L..41B}, \citet{2008ApJ...685.1063B} and \citet{2011ApJ...727...29M} estimated the rotational energies of the BHs at the time they were born and estimated whether a Blandford-Znajek \citep[BZ;][]{1977MNRAS.179..433B} central engine may have triggered a lGRB in any of these BHBs.
In \citet{2008ApJ...685.1063B} it was found that, most likely, LMC X$-$3 is a likely lGRB relic, unlike most Galactic BHBs where too much rotational energy may have destroyed the BZ engines too soon.
In \citet{2008ApJ...689L...9M} and \citet{2011MNRAS.413..183M} it was shown that the large spins observed in some Galactic high-mass X-ray binaries (HMXBs) cannot be natal but they rather have to be acquired post-BH formation.
Thus, these HMXBs are not good candidates for relics of lGRBs. \\
\citet{2012ApJ...752...32W} have suggested that this binary channel may not produce lGRBs as the internal magnetic fields and/or dynamos may extract the angular momentum off the stellar core.  \citet{2014ApJ...781....3M} shows that this diffuculty may be overcome if the internal field is low enough and a magnetar-like field can be produced during core collapse.\\
Now, LMXBs (low-mass XBs), which have too much rotational energy may initially trigger an internal jet which quickly dies after the BZ engine is destroyed as a SN is launched.
Also, many succesful lGRBs may not be pointed in our direction.
Similarly, were some stars to develop a rapidly rotating core without losing their envelope, an internal jet could be launched which would be quenched before reaching the stellar surface.
All of these should produce a strong thermal neutrino signal at the central engine.
These neutrinos should be observable within a several-Mpc range.\\
%Here we study some of the characteristics of these neutrino pulses.
The  properties of neutrinos get modified when they propagate in a strongly magnetized medium.
Depending on the flavor of the neutrinos, they interact with different effective potentials because electron neutrinos ($\nu_e$) interact with electrons via both, neutral and charged currents (CC), whereas muon and tau neutrinos ($\nu_\mu$ and $\nu_\tau$) interact with electrons only via the neutral current (NC).  
This induces a coherent effect in which maximal conversion of $\nu_e$ into $\nu_\mu$ ($\nu_\tau$) takes place even for a small, intrinsic, mixing angle.  
The resonant conversion of neutrino from one flavor to another due to the medium effect, is well known as the Mikheyev-Smirnov-Wolfenstein effect. %%%%%\citep{wol78}.
In this work, we study  the propagation and resonant oscillation of thermal neutrinos in both, a weakly- and  highly-magnetized fireball.  We take into account  the two-neutrino mixing (solar, atmospheric and accelerator parameters). Finally, we discuss our results in the GRB framework.

\section{Neutrino  Effective Potential}\label{sec-Justification}

We use the finite-temperature, field-theory formalism to study the effect of a heat bath on the propagation of elementary particles \citep{2009PhRvD..80c3009S,2007MPLA...22..213G, 1990PhRvD..42.4123N}. 
The effect of the magnetic field is taken into account through Schwinger's proper-time method \citep{1951PhRv...82..664S}. %%%%\citep{sch51}.  
The effective potential of a particle is calculated from the real part of its self-energy diagram. 
The neutrino field equation of motion in a magnetized medium is,
\be
[ {\rlap /k} -\Sigma(k) ] \Psi_L=0,
\label{disneu}
\ee
where the neutrino self-energy operator $\Sigma(k)$ is a Lorentz scalar which depends on the characterized parameters of  the medium, as for instance, chemical potential, particle density, temperature, magnetic field, etc.  Solving this equation and using Dirac's algebra we can write the dispersion relation $V_{eff}=k_0-|{\bf k}|$ as a function of Lorentz scalars: 
\be
V_{eff}=b-c\,\cos\phi-a_{\perp}|{\bf k}|\sin^2\phi,
\label{poteff}
\ee
where $\phi$ is the angle between the neutrino momentum and the magnetic field vector.   
The Lorentz scalars $a$, $b$ and $c$ are functions of the neutrino energy, momentum and the magnetic field.
They can be calculated from the neutrino self-energy due to charge-current and neutral-current interaction with the background particles.
%%%%%%%%%%%%%%%%%%%%%%%%%%%%%%%%%%%%%%%%%%%%%%%%%%%%%%%%%%%%%%%%%%%%%%%%%%%
%%%%%%%%%%%%%%%%%%%%%%%%%%%%%%%%%%%%%%%%%%%%%%%%%%%%%%%%%%%%%%%%%%%%%%%%%%%

\subsection{One-loop  neutrino self-energy}

The total one-loop  neutrino self-energy  in a magnetized medium is given by %\citep{bra07, eli04,erd98,sah09a, sah09b},
\be
\Sigma(k) = \Sigma_W(k) + \Sigma_Z (k)+ \Sigma_t (k)
\label{tsen}
\ee
where $\Sigma_W (k)$ is the  $W$-exchange,  $\Sigma_Z (k)$ is the $Z$-exchange and   $\Sigma_t (k)$ represents  the tadpole.  The W-exchange diagram to the one-loop self-energy is 

\bary
-i\Sigma_W(k)&=&R\biggl[\int\frac{d^4p}{(2\pi)^4}\left(\frac{-ig}{\sqrt{2}}\right)
\gamma_\mu  iS_{\ell}(p)\cr
&& \left(\frac{-ig}{\sqrt{2}}\right)\gamma_\nu   i W^{\mu \nu}(q)\biggr]L\,
\label{Wexch}
\eary
where $g^2=4\sqrt2 G_FM_W^2$  is the weak coupling constant, $W^{\mu\nu}$ depicts the W-boson propagator which, in the eB$\ll$ M$^2_W$ limit and in unitary gauge, is given by %\citep{erd98,sah09b},
\be
W^{\mu\nu}(q)=\frac{g^{\mu\nu}}{M^2_W}\biggl(1+\frac{q^2}{M^2_W}  \biggr)-\frac{q^\mu q^\nu}{M^4_W}+\frac{3ie}{2M^4_W}F^{\mu\nu},
\ee
here M$_W$ is the W-boson mass, $g^{\mu\nu}$ is the  metric tensor, $F^{\mu\nu}$ is the electromagnetic field tensor and $e$ being the magnitude of the electron charge.  $S_\ell(p)$ stands for the charged lepton propagator which can  be separated into two charged propagators; one in presence of a uniform background magnetic field ($S^0_\ell(p)$) and another in a magnetized medium ($S^\beta_\ell (p)$). 
It can be written as, 
\be
S_\ell (p) = S^0_\ell(p) + S^\beta_\ell (p)\,.
\label{slp}
\ee
Assuming that the z-axis points in the direction of the magnetic field ${\rm B}$, we can express the charged lepton propagator  in presence of a uniform background magnetic field  as,
\be
i S^0_\ell(p) = \int_0^\infty e^{\phi(p,s)} G(p,s)\,ds\,,
\label{sl0p}
\ee
where  the functions $\phi(p,s)$ and $G(p,s)$ are give by,
\begin{eqnarray}
G(p,s)&=&\sec^2 z \left[{\rlap A/} + i {\rlap B/} \;\gamma_5\right. \nonumber \\
      &+& \left. m_\ell(\cos^2 z - i \Sigma^3 \sin z \cos z)\right]\nonumber \\%\cr
\phi(p,s)&=&is(p_0^2 - m_\ell^2) - is[p^2_3 + \frac{\tan z}{z} p^2_\perp]\,,
\label{phaselu}
\end{eqnarray}
here $m_\ell$ is the mass of the charged lepton, $p^2_\parallel = p_0^2 - p_3^2$ and  $p^2_\perp = p_1^2 + p_2^2$ are the projections  of the momentum on the magnetic field direction  and  $z= e{ B}s$, 
Additionally,  the covariant  vectors are given as follows, 
\[
A_\mu = p_\mu -\sin^2 z (p\cdot u\,\, u_\mu - p\cdot b \,\,b_\mu)\, ,
\] 
\[
B_\mu = \sin z\cos z (p\cdot u \,\,b_\mu - p\cdot b \,\,u_\mu)\,,
\] 
and 
\[
\Sigma^3 = \gamma_5 {\rlap /b} {\rlap /u}\,.
\]
The charged lepton propagator in a magnetized medium is given by, 
\be
S^\beta_\ell(p) = i \eta_F(p\cdot u)\int_{-\infty}^\infty e^{\phi(p,s)} G(p,s)
\,ds\,,
\label{slbp}
\ee
where $\eta_F(p\cdot u)$ contains the distribution functions of the particles in the medium which are given by:
\be
\eta_F (p \cdot u) = \frac{\theta(p\cdot u)}{e^{\beta(p\cdot u -
\mu_\ell)} + 1 } +
\frac{\theta(- p\cdot u)}{e^{-\beta(p\cdot u - \mu_\ell)} + 1}\,,
\label{eta}
\ee
where $\beta$ and $\mu_\ell$ are the inverse of the medium temperature and the chemical potential of the charged lepton.

The Z-exchange diagram to the one-loop self-energy is 
%{\scriptsize 
\bary
-i\Sigma_Z(k)&=&R\,\biggl[\int\frac{d^4 p}{(2\pi)^4}\left(\frac{-ig} {\sqrt{2}\cos\theta_W}\right) 
\gamma_\mu \,iS_{\nu_\ell}(p)\cr
&&\left(\frac{-ig}{\sqrt{2}\cos\theta_W}\right) \gamma_\nu\,i Z^{\mu \nu}(q)\biggr]\,L\,,
\label{Zexch}
\eary
%}
%
$\theta_W$ is the Weinberg angle, Z$^{\mu\nu}$(q) is the Z-boson propagator in vaccum, S$_{\nu_l}$ is the neutrino propagator in a thermal bath of neutrinos. \\
The Tadpole diagram to the one-loop self-energy is 
%{\scriptsize 
\bary
i\Sigma_t(k)&=&R\,\biggl[ \left(\frac{g}{2\cos \theta_W}\right)^2\,
\gamma_\mu\,iZ^{\mu \nu}(0)\int\frac{d^4 p}{(2\pi)^4} \cr
&& {\rm Tr} \left[\gamma_\nu \,(C_V + C_A \gamma_5)\,iS_{\ell}(p)\right]\,\biggr]L\,,
\label{tad}
\eary
%}
%
where the quantities $C_V$ and $C_A$ are the vector and axial-vector coupling constants which come in the neutral-current interaction of
electrons, protons ($p$), neutrons ($n$) and neutrinos with the $Z$ boson.  Their forms are as follows,
\bary
C_V=\left \{\begin{array}
{r@{\quad\quad}l}
-\frac{1}{2}+2\sin^2\theta_W & e\\
\frac{1}{2} & {\nu}\\ \frac{1}{2}-2\sin^2\theta_W & {{p}}\\
-\frac{1}{2} & {{n}}
\end{array}\right.,
\label{cv}
\eary
and
\bary
C_A=\left \{\begin{array}
{r@{\quad\quad}l}
-\frac{1}{2} & {\nu},{{p}}\\
\frac{1}{2} & e,{{n}}
\end{array}\right..
\label{ca}
\eary
%%%%%%%
%%%%%%%
By evaluating  each contribution of one-loop neutrino energy, the effective potential in the    weak \citep{2009JCAP...11..024S} ($B \ll m^2_e/e$) and strong ($B \gg m^2_e/e$) \citep{2014arXiv1401.1581F,2014arXiv1401.1908F,2014arXiv1401.3787F}  -field limit  is given respectively by
{\small
\bary
V_{eff,w}&=&\frac{\sqrt2 G_F\,m_e^3}{\pi^2}\frac{B}{B_c}\biggl[\sum^\infty_{l=0} (-1)^l \sinh\alpha\,K_1(\sigma) \biggl\{ 1+C_{V_e}\cr
&&+ \frac{m_e^2}{m^2_W}\biggl(\frac32+2\frac{E^2_\nu}{m^2_e} +\frac{B}{B_c}  \biggr)-\biggl(1-C_{Ae}+ \frac{m_e^2}{m^2_W}\cr
&&\times \biggl(\frac12-2\frac{E^2_\nu}{m^2_e} +\frac{B}{B_c}  \biggr) \biggr)\cos\phi  \biggr\}-4\frac{m_e^2}{m^2_W}\frac{E_\nu}{m_e}\sum^\infty_{l=0} (-1)^l\cr
&&\times \cosh\alpha\biggl\{ \frac34K_0(\sigma)+\frac{K_1(\sigma)}{\sigma}-\frac{K_1(\sigma)}{\sigma}\cos\phi\biggr\}\biggr]
\label{poteff1}  %%%EMM, 7/6: Check ref on this label
\eary
}
and
{\small
\bary
V_{eff,s}&=& \sqrt2 G_F \frac{m^3_e}{\pi^2}  \biggl[ \sum_{l=0}^{\infty} (-1)^l \sinh\, \alpha \biggl[\biggl(1+\frac32\frac{m_e^2}{M_W^2} -\frac{eB}{M_W^2}\biggr) \cr
&&\biggl( \frac{2}{\sigma} K_2(\sigma) - \frac{B}{B_c} K_1(\sigma) \biggr)\biggr]-\frac{B}{B_c}\biggl(1+\frac{m_e^2}{2 M_W^2}-\frac{eB}{M_W^2} \biggr )\cr
&&\times K_1(\sigma)-\frac{2 m_e E_{\nu}}{M^2_W}   \sum_{l=0}^{\infty} (-1)^l\cosh\, \alpha \biggl[\biggl(\frac{8}{\sigma^2}-\frac52 \frac{B}{B_c} \biggr )\cr
&&\times K_0(\sigma)+\biggl( 2-4 \frac{B}{B_c} + \frac{16}{\sigma^2} \biggr )\frac{K_1(\sigma)}{\sigma} \biggr ]\biggr],
\eary
}
where,
\be
\alpha=\beta\mu(l+1)\hspace{0.5cm} {\rm and}\hspace{0.5cm} \sigma=\beta m_e(l+1).
\ee

%\begin{widetext}
%\bary
%V_{eff,s}&=& \sqrt2 G_F \frac{m^3_e}{\pi^2}  \biggl [ \sum_{l=0}^{\infty} (-1)^l \sinh\, \alpha  \biggl[\biggl(1+\frac32\frac{m_e^2}{M_W^2} -\frac{eB}{M_W^2}\biggr) \biggl( \frac{2}{\sigma} K_2(\sigma) - \frac{B}{B_c} K_1(\sigma) \biggr)-\frac{B}{B_c}\biggl(1+\frac{m_e^2}{2 M_W^2}-\frac{eB}{M_W^2} \biggr ) K_1(\sigma)\cr
%&-&\frac{2 m_e E_{\nu}}{M^2_W}   \sum_{l=0}^{\infty} (-1)^l\cosh\, \alpha \biggl[\biggl(\frac{8}{\sigma^2}-\frac52 \frac{B}{B_c} \biggr )K_0(\sigma)+\biggl( 2-4 \frac{B}{B_c} + \frac{16}{\sigma^2} \biggr )\frac{K_1(\sigma)}{\sigma}  \biggr ],
%\eary
%\end{widetext}

\section{Two-Neutrino Mixing}

Here we consider the neutrino oscillation process $\nu_e\leftrightarrow \nu_{\mu, \tau}$. The evolution equation for the propagation of neutrinos in the above medium is given by
\be
i
{\pmatrix {\dot{\nu}_{e} \cr \dot{\nu}_{\mu}\cr}}
={\pmatrix
{V_{eff}-\Delta \cos 2\theta & \frac{\Delta}{2}\sin 2\theta \cr
\frac{\Delta}{2}\sin 2\theta  & 0\cr}}
{\pmatrix
{\nu_{e} \cr \nu_{\mu}\cr}},
\ee
where $\Delta=\delta m^2/2E_{\nu}$, $V_{eff}$ is the potential difference between $V_{\nu_e}$ and $V_{\nu_{\mu, \tau}}$,    $E_{\nu}$ is the neutrino energy and $\theta$ is the neutrino mixing angle. The conversion probability for the above process at a given time $t$ is given by
\be
P_{\nu_e\rightarrow {\nu_{\mu}{(\nu_\tau)}}}(t) = 
\frac{\Delta^2 \sin^2 2\theta}{\omega^2}\sin^2\left (\frac{\omega t}{2}\right
),
\label{prob}
\ee
with
\be
\omega=\sqrt{(V_{eff}-\Delta \cos 2\theta)^2+\Delta^2 \sin^2
    2\theta}.
\ee
The potential for the above oscillation process is given by eq. (\ref{poteff}). The oscillation length for the neutrino is given by
\be
L_{osc}=\frac{L_v}{\sqrt{\cos^2 2\theta (1-\frac{V_{eff}}{\Delta \cos 2\theta}
    )^2+\sin^2 2\theta}},
\label{osclength}
\ee
and the resonance condition can be written as 
\be
V_{eff} -  \frac{\delta m^2}{2E_{\nu}} \cos 2\theta = 0.
\label{reso}
\ee
Following \citet{2014MNRAS.437.2187F} we apply the neutrino oscillation parameters in the resonance condition (eq. \ref{reso}). 
\begin{figure}
\vspace{0.3cm}
{ \centering
\resizebox*{0.45\textwidth}{0.27\textheight}
{\includegraphics{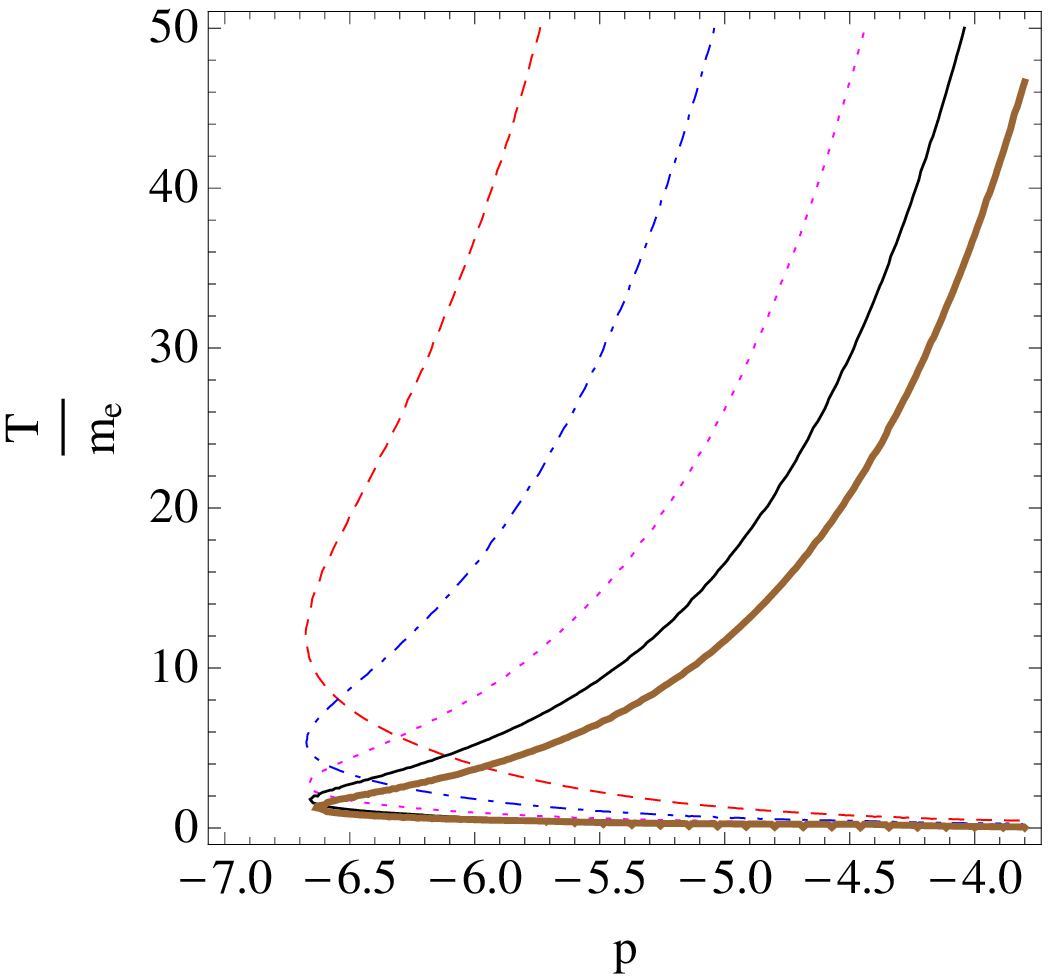}}
\resizebox*{0.45\textwidth}{0.27\textheight}
{\includegraphics{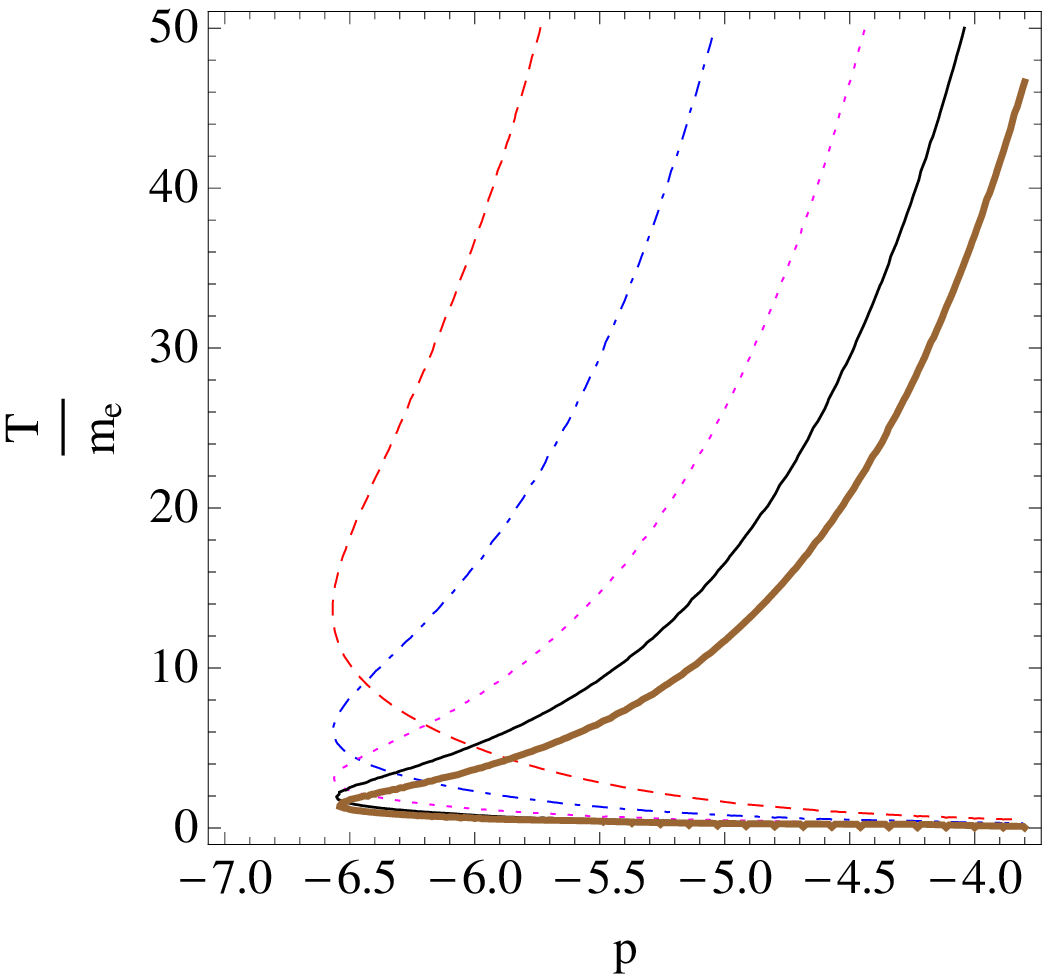}}
\resizebox*{0.45\textwidth}{0.27\textheight}
{\includegraphics{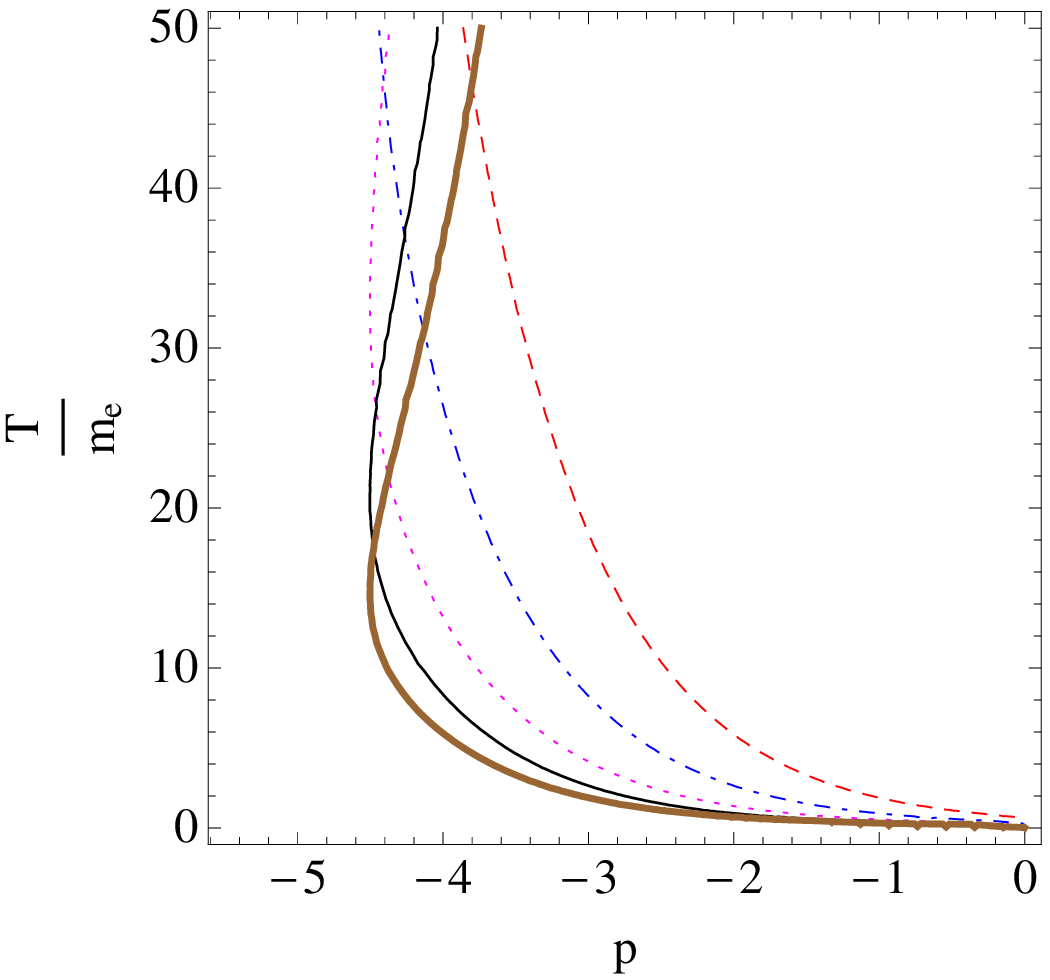}}
}
\caption{ Plot of temperature  ($T/m_e$) as a function of chemical potential  ($\mu=m_e10^p$) for which the resonance condition is satisfied.  We have used the best parameters of the two-flavor  solar (top),  atmospheric (middle) and  accelerator (bottom) neutrino oscillation, B=0.1 B$_c$ and taken five  different neutrino energies:  E$_\nu=$1 MeV (red dashed line),  E$_\nu=$5 MeV (blue dot-dashed line), E$_\nu=$20 MeV (magenta dotted line), E$_\nu=$50 MeV (black thin-solid line) and E$_\nu=$100 MeV (brown thick-solid line).}
\label{ptimew}
\end{figure}

\begin{figure}
\vspace{0.3cm}
{ \centering
\resizebox*{0.45\textwidth}{0.27\textheight}
{\includegraphics{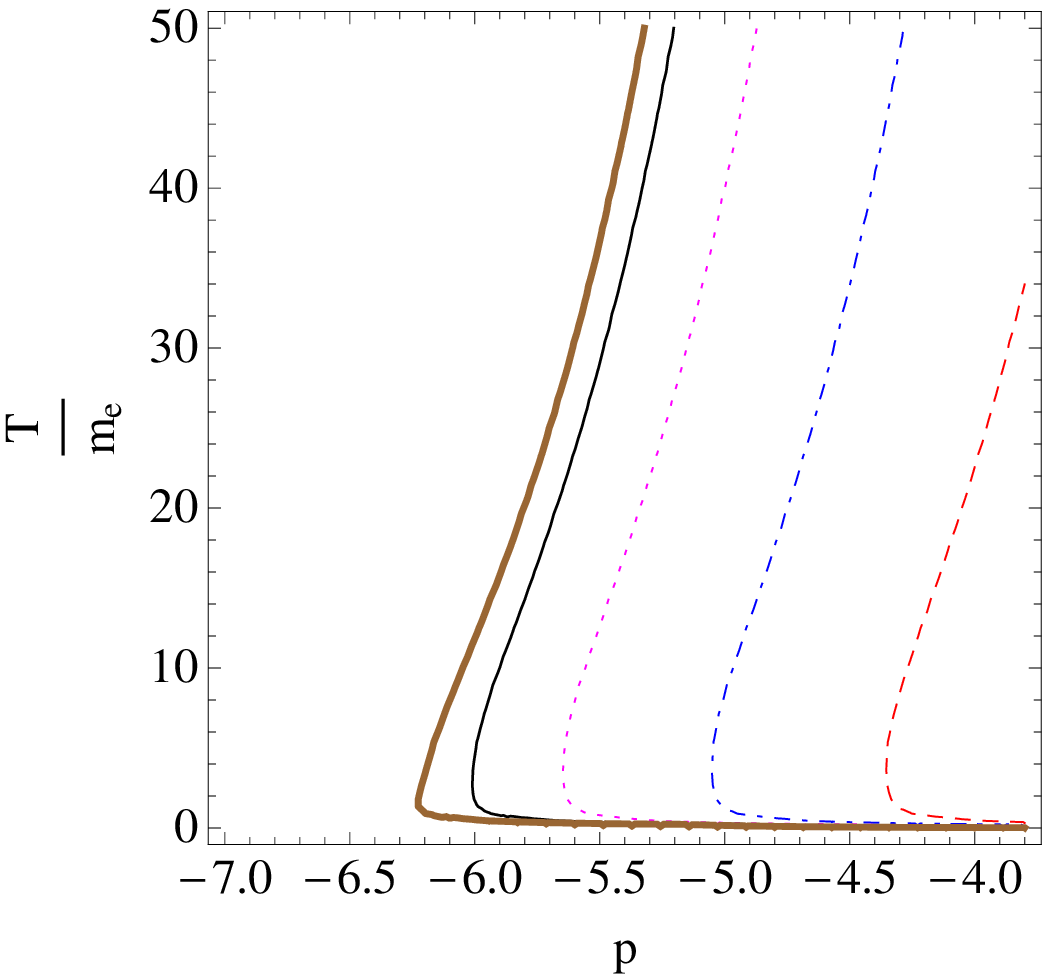}}
\resizebox*{0.45\textwidth}{0.27\textheight}
{\includegraphics{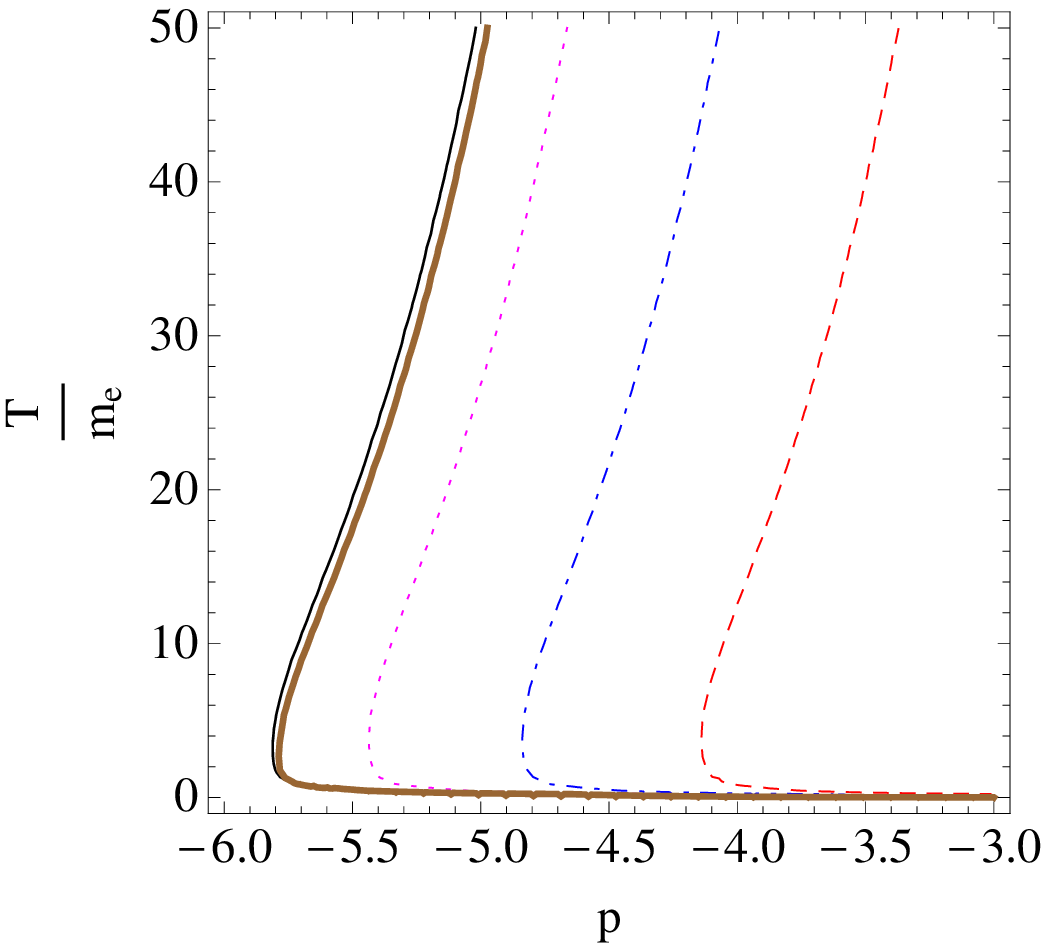}}
\resizebox*{0.45\textwidth}{0.27\textheight}
{\includegraphics{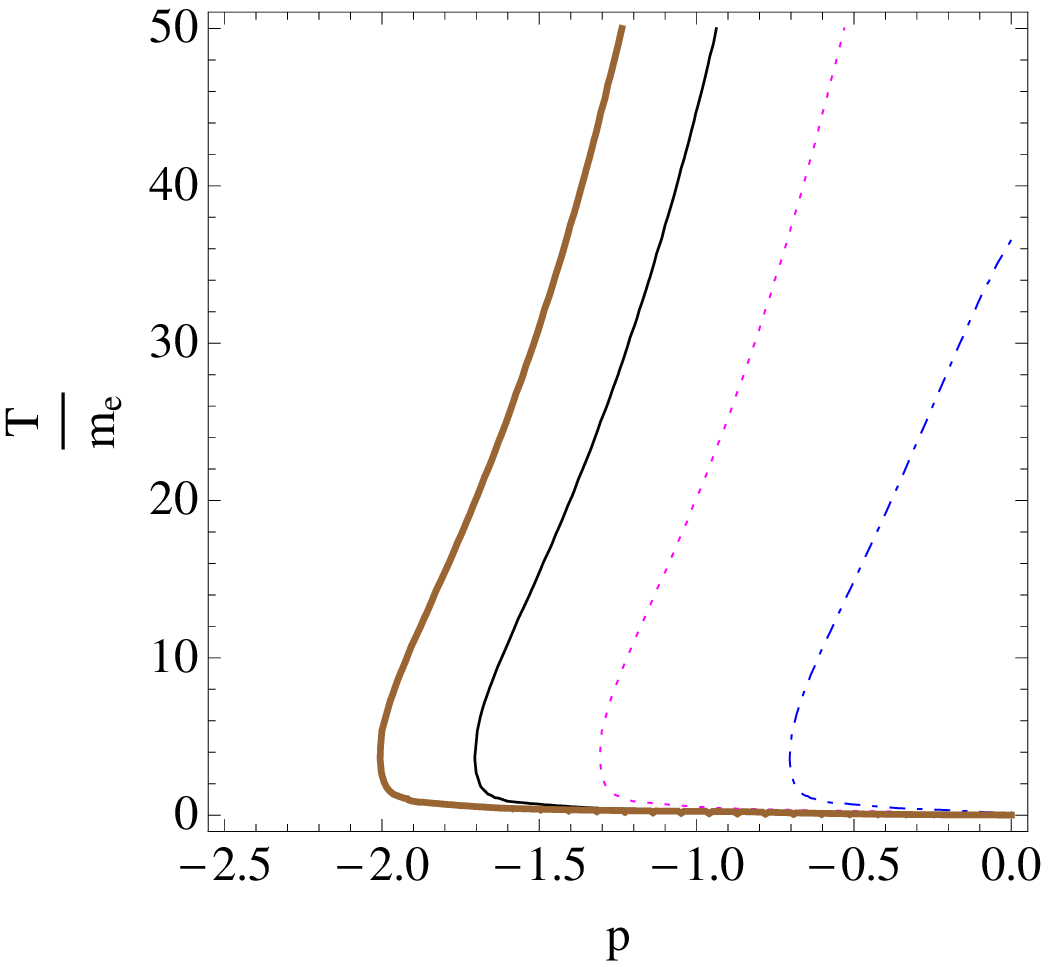}}
}
\caption{Plot of temperature  ($T/m_e$) as a function of chemical potential  ($\mu=m_e10^p$) for which the resonance condition is satisfied.  We have used the best parameters of the two-flavor  solar (top),  atmospheric (middle) and  accelerator (bottom) neutrino oscillation, B=10 B$_c$ and taken five  different neutrino energies:  E$_\nu=$1 MeV (red dashed line),  E$_\nu=$5 MeV (blue dot-dashed line), E$_\nu=$20 MeV (magenta dotted line), E$_\nu=$50 MeV (black thin-solid line) and E$_\nu=$100 MeV (brown thick-solid line). }
\label{ptimes}
\end{figure}
%%%%% Falta citar figuras en texto!!!

\section{Results and Conclusions}

We have plotted the resonance condition for  neutrino oscillation in weakly and strongly magnetized collapsars at the base of the jet when the  temperature is in the range of 1- 25 MeV. We have used  effective potential $V_{eff}$ up to order M$_W^4$ in the strong- and weak-field limit, the best parameters for the two- solar \citep{2011arXiv1109.0763S}, atmospheric \citep{2011PhRvL.107x1801A} and accelerator \citep{1996PhRvL..77.3082A, 1998PhRvL..81.1774A} neutrinos and  the neutrino energies ($E_\nu = 1$, 5, 20, 50 and 100 MeV). The analysis of resonance condition shows that,  the temperature as a function of chemical potential is degenerate  for the weak limit and not in the strong limit.  In both cases, the chemical potential is biggest when accelerator parameters  are used (19 eV - 50 keV and 5 keV - 0.15 MeV, respectively )  and smallest for solar parameters (0.08 eV - 5 eV and 0.25 - 50 eV, respectively).\\
The baryon load, resonance length and leptonic symmetry (two and three flavors) will be estimated and studied in Fraija \& Moreno M\'endez in progress.
\acknowledgments
NF gratefully acknowledges a Luc Binette-Fundaci\'on UNAM Posdoctoral Fellowship. EMM was supported by a CONACyT fellowship and projects CB-2007/83254 and CB-2008/101958.  This research has made use of NASAs Astrophysics Data System as well as arXiv. 
%

%\bibliographystyle{apj}
%\bibliography{bibliography}

\end{document}